\documentclass[a4paper,twocolumn]{jpconf}

\usepackage{graphicx}
\usepackage{subfigure}

\begin{document}
\twocolumn[
\title{Precise measurement of HFS of positronium}

\author{A Ishida$^1$, G Akimoto$^1$, K Kato$^1$, T Suehara$^1$, T Namba$^1$, S Asai$^1$, \\
T Kobayashi$^1$, H Saito$^2$, M Yoshida$^3$, K Tanaka$^3$, A Yamamoto$^3$, \\
I Ogawa$^4$, S Kobayashi$^4$ and T Idehara$^4$}

\address{$^1$ Department of Physics and ICEPP, the University of Tokyo, 7-3-1 Hongo, Bunkyo-ku, Tokyo, 113-0033, Japan}
\address{$^2$ Institute of Physics, the University of Tokyo, 3-8-1 Komaba, Meguro-ku, Tokyo, 153-8902, Japan}
\address{$^3$ High Energy Accelerator Research Organization (KEK), 1-1 Oho, Tsukuba, Ibaraki, 305-0801, Japan}
\address{$^4$ FIR Center, University of Fukui, 3-9-1 Bunkyo, Fukui, 910-8507, Japan}

\ead{ishida@icepp.s.u-tokyo.ac.jp}

\begin{abstract}
The ground state hyperfine splitting in positronium, $\Delta _{\mathrm{HFS}}$, is sensitive to high order corrections of QED. A new calculation up to $O(\alpha ^3)$ has revealed a $3.9\,\sigma$ discrepancy between the QED prediction and the experimental results. This discrepancy might either be due to systematic problems in the previous experiments or to contributions beyond the Standard Model. We propose an experiment to measure $\Delta _{\mathrm{HFS}}$ employing new methods designed to remedy the systematic errors which may have affected the previous experiments. Our experiment will provide an independent check of the discrepancy. The measurement is in progress and a preliminary result of $\Delta _{\mathrm{HFS}} = 203.399 \pm 0.029\,\mathrm{GHz}\,(143\,\mathrm{ppm})$ has been obtained. A measurement with a precision of $O(1)$\,ppm is expected within a few years.
\end{abstract}
]
\section{Introduction}
Positronium (Ps), a bound state of an electron and a positron, is a purely leptonic system 
which allows for very sensitive tests of QED. 
The precise measurement of the hyperfine splitting between orthopositronium (o-Ps, 1$^3S_1$) and parapositronium (p-Ps, 1$^1S_0$) (Ps-HFS) provides a good test of bound state QED. 
Ps-HFS is expected to be relatively large (for example compared to hydrogen HFS) due to a relatively large 
spin-spin interaction, and also due to the contribution from vacuum oscillation 
(o-Ps $ \rightarrow \gamma ^{\ast} \rightarrow$ o-Ps). 
The contribution from vacuum oscillation is sensitive to new physics beyond the Standard Model. 

\begin{figure}
\begin{center}
\includegraphics[width=0.48\textwidth]{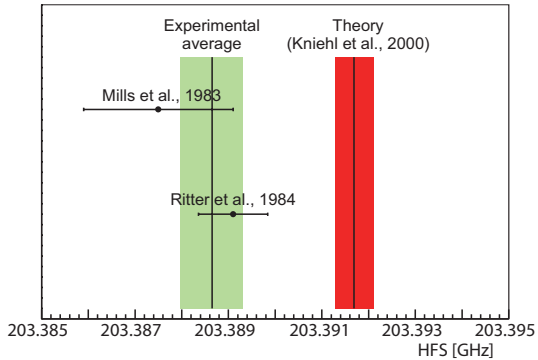}
\caption{\label{fig:history}Measured and theoretical values of Ps-HFS.}
\end{center}
\end{figure}

Figure \ref{fig:history} shows the measured and theoretical values of Ps-HFS. 
The combined value from the results of the previous 2 experiments is 
$\Delta _{\mathrm{HFS, exp}} = 203.388\,65(67)\,\mathrm{GHz} \,(3.3\,\mathrm{ppm})$\\~\cite{hughes,mills}. 
Recent developments in nonrelativistic QED (NRQED) have added 
$O(\alpha ^3)$ corrections to the theoretical prediction which now stands at 
$\Delta_{\mathrm{HFS, th}} = 203.391\,69(41)\,\mathrm{GHz} \\ 
\,(2.0\,\mathrm{ppm})$~\cite{kniehl}.
The discrepancy of \\ 3.04(79)\,MHz (15\,ppm, 3.9\,$\sigma$) between \\$\Delta _{\mathrm{HFS, exp}}$ 
and $\Delta_{\mathrm{HFS, th}}$ might either be due to the common systematic uncertainties in the previous experiments or to 
new physics beyond the Standard Model. 

There are two possible common systematic uncertainties in the previous experiments. 
One is the unthermalized o-Ps contribution which results in an underestimation of the material effect. 
This effect has already been shown to be significant~\cite{kataoka}.
The other is the uncertainty in the magnetic field uniformity which was cited as 
the most significant systematic error by previous experimenters. 

\section{Experimental setup}
\begin{figure}
\begin{center}
\includegraphics[width=0.48\textwidth]{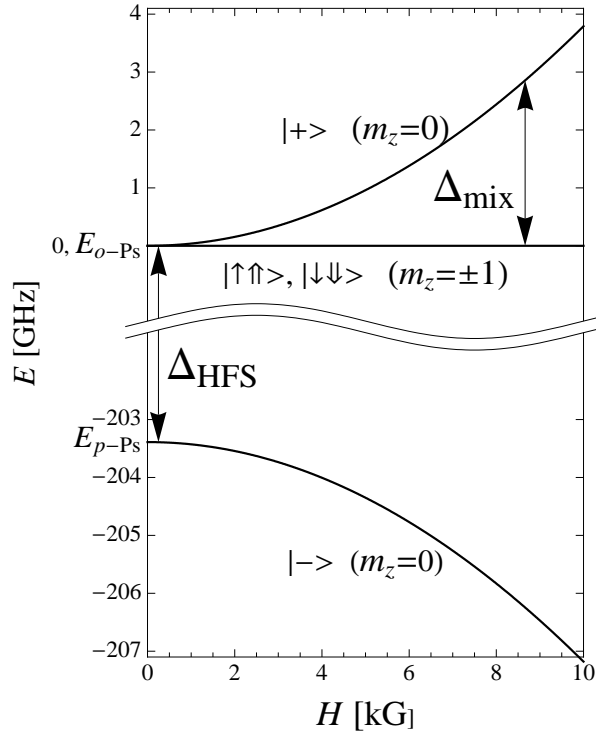}
\caption{\label{fig:pslevel}Zeeman energy levels of Ps in its ground state.}
\end{center}
\end{figure}

The energy levels of the ground state of Ps are shown as a function of static magnetic field 
in Figure \ref{fig:pslevel}. Due to technical difficulties in directly stimulating 
$\Delta _{\mathrm{HFS}}$, we make an indirect measurement by stimulating the 
transition $\Delta _{\mathrm{mix}}$. This is the same approach as previous experiments. 
The relationship between $\Delta _{\mathrm{HFS}}$ and $\Delta _{\mathrm{mix}}$ is 
given by the Breit-Rabi equation 
\begin{equation}
\Delta _{\mathrm{mix}} = \frac{1}{2} \Delta _{\mathrm{HFS}} \left( \sqrt{1+x^2} - 1 \right) \, ,
\label{eq:Zeeman}
\end{equation}
in which $x=2g^{\prime}\mu _B H  / h\Delta _{\mathrm{HFS}}$. $g^{\prime} = g\left( 1-\frac{5}{24} \alpha ^2 \right)$ 
is the $g$ factor for a positron (electron) in Ps~\cite{gfactor}, 
$\mu _B$ is the Bohr magneton, $H$ is the static magnetic field, and $h$ is the Plank constant.

In a static magnetic field, the p-Ps state mixes with the $m_z=0$ substate of o-Ps 
hence the latter state annihilates into 2 $\gamma$-rays with a lifetime of about 
8\,ns (with our experimental conditions). 
The $m_z = \pm 1$ substates of o-Ps annihilate into 3 $\gamma$-rays with a lifetime of 
about 140\,ns. When a microwave field with a frequency of $\Delta _{\mathrm{mix}}$ is 
applied, transitions between the $m_z = 0$ and $m_z = \pm 1$ substates of o-Ps are 
induced so that the 2 $\gamma$-ray annihilation rate increases. 
This increase is our experimental signal. 
 
Following are the main improvements in our experiment which we expect will significantly reduce 
systematic errors present in previous experiments. 

\begin{figure}
\begin{center}
\subfigure[The large bore superconducting magnet and the microwave waveguide. Microwaves are guided through the waveguide into the cavity.]{
	\includegraphics[width=0.48\textwidth]{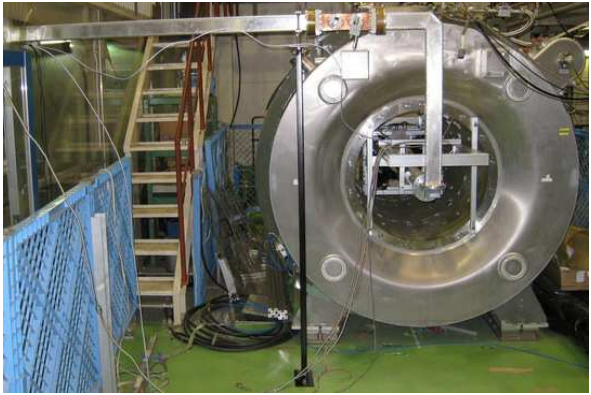}
	\label{fig:apparatus1}}
\subfigure[Inside of the magnet. The $\beta$-tagging system, the $\gamma$-ray detectors and the microwave cavity are located at the center of the magnet.]{
	\includegraphics[width=0.48\textwidth]{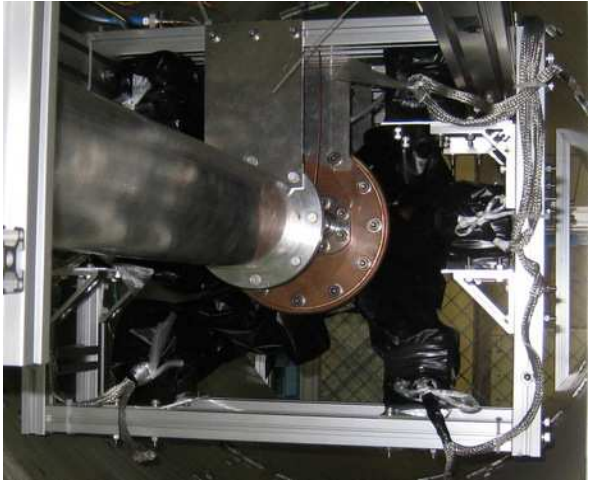}
	\label{fig:apparatus2}}
	\caption{Photographs of the experimental setup.}
	\label{fig:apparatus}
\end{center}
\end{figure}

\subsection{Large bore superconducting magnet (Figure \ref{fig:apparatus})}
A large bore superconducting magnet is 
used to produce the magnetic field which induces the Zeeman splitting ($\sim 0.866\,\mathrm{T}$). 
The bore diameter of the magnet is 800\,mm, and its length is 2\,m. 
The large bore diameter means that there is good uniformity in the magnetic field in the 
region where Ps is formed (70\,ppm at 0.61\,G without utilization of any compensation). 

Furthermore, the magnet is operated in persistent current mode, making the stability of the magnetic field 
better than 1\,ppm.

\subsection{$\beta$-tagging system and timing information}
The positron source is 19\,$\mu$Ci (700\,kBq) of $ \mathrm{^{22}Na}$. 
A plastic scintillator 10\,mm in diameter and 0.2\,mm thick is used to tag positrons emitted from the $\mathrm{^{22}Na}$.
The scintillation light is detected by photomultipliers 
and provides a start signal which corresponds to the time of Ps formation. 

The positron then enters the microwave cavity, forming Ps in the $\mathrm{N_2}$ gas contained therein. 

Ps decays into photons that are detected with LaBr$_3$\,(Ce) scintillators. 
Accumulating measurements of the times of positron emission and $\gamma$-detection results in 
decay curves of Ps as shown in Figure \ref{fig:timing_spectra}.
The timing information is used to improve the accuracy of the measurement of $\Delta _{\mathrm{HFS}}$ as follows: 
\begin{enumerate}
\item Imposing a time cut means that we can select well thermalized Ps, reducing the unthermalized o-Ps contribution. 
It should also be possible to precisely measure the contributions of 
unthermalized o-Ps, and of material effects (we plan to make such measurements in future runs). 
\item A time cut also allows us to avoid the prompt peak (contributions of 
simple annihilation and of fast p-Ps decay), which greatly increases the S/N of the measurement. 
\end{enumerate}

\subsection{High performance $\gamma$-ray detectors}
Six $\gamma$-ray detectors are located around the microwave cavity to detect 
the 511\,keV annihilation $\gamma$-rays. 
LaBr$_3$\,(Ce) scintillators, 1.5 inches in diameter and 2 inches long are used. 
LaBr$_3$\,(Ce) scintillators have 
good energy resolution (4\,\% FWHM at 511\,keV) and timing resolution (0.2\,ns FWHM at 511\,keV), 
and have a short decay constant (25.6\,ns). 
The good energy resolution and the high counting rate of LaBr$_3$ results in very good overall performance 
for measuring 2$\gamma$ decays. 
In particular the good energy resolution allows us to efficiently separate 2$\gamma$ events from 
3$\gamma$ events, negating the need to use a back-to-back geometry to select 2$\gamma$ events, thus 
greatly increasing the acceptance of our setup. 

This $\gamma$-ray detector system greatly reduces the statistical error in the measurement. 
\subsection{RF system}

Microwaves are produced by 
a local oscillator signal generator and amplified to 500\,W with a 
GaN amplifier (R\&K A2856BW200-5057-R).
 
The microwave cavity is made with oxgen-free copper; 
the inside of the cavity is a cylinder 128\,mm in diameter and 
100\,mm long. The side wall of the cavity is only 2\,mm thick in order to 
allow the $\gamma$-rays to efficiently escape. 
The cavity is operated in the TM$_{110}$ mode. The resonant frequency is 2.8532\,GHz and 
$Q_L = 14700 \pm 50$. The cavity is filled with 1.5\,atm gas 
(90\,\% N$_2$ and 10\,\% iso-C$_4$H$_{10}$) with a gas-handling system. Iso-C$_4$H$_{10}$ is used as the quenching gas 
to remove background 2 $\gamma$-ray annihilations.

\section{Analysis and current status}

The first run of the experiment is ongoing 
using the large bore magnet with no compensation (compensation magnets to reduce 
the uniformity to $O(1)$\,ppm are planned but are not yet installed). 
The 2 $\gamma$-ray annihilation rate has been measured at various magnetic field strengths with a fixed rf frequency.

\subsection{Data analysis}

\begin{figure}
\begin{center}
\includegraphics[width=0.48\textwidth]{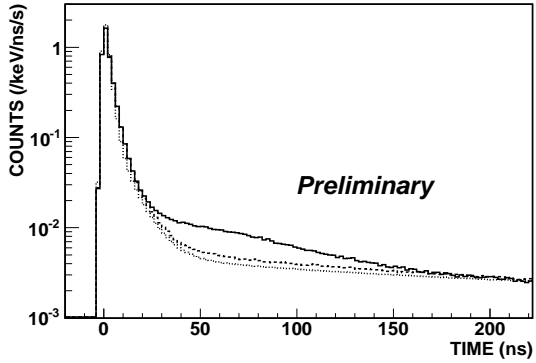}
\caption{\label{fig:timing_spectra}Decay curves of Ps at different magnetic field strengths. The solid line is 0.8659\,T RF\,450\,W, the dashed line is 0.8614\,T RF\,450\,W, and the dotted line is RF\,OFF. An energy window of 492--530\,keV is applied.}
\end{center}
\end{figure}

\begin{figure}
\begin{center}
\includegraphics[width=0.48\textwidth]{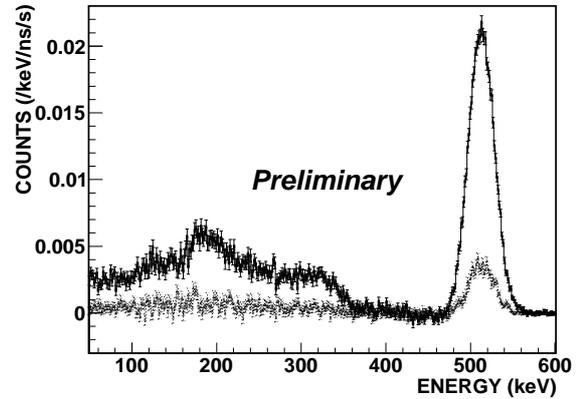}
\caption{\label{fig:energy_spectra}Energy spectra at different magnetic field strengths. The solid line is 0.8659\,T RF\,450\,W and the dashed line is 0.8614\,T RF\,450\,W. A timing window of 30--200\,ns is applied, and the accidental spectra from 700--900\,ns and RF\,OFF have been subtracted.}
\end{center}
\end{figure}

\begin{figure}
\begin{center}
\includegraphics[width=0.48\textwidth]{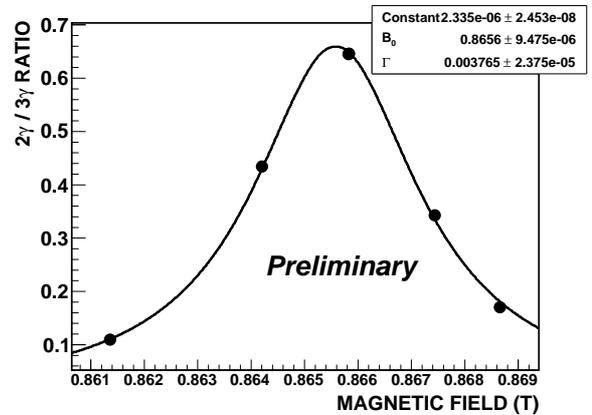}
\caption{\label{fig:resonance_line}Resonance line at 1.5\,atm. The data are fitted with the Breit-Wigner function (Equation (\ref{bw})). Error bars are smaller than the marker size.}
\end{center}
\end{figure}

Figure \ref{fig:timing_spectra} shows examples of measured timing spectra. The peak coming from prompt annihilation and 
p-Ps decay is followed by the decay curve of o-Ps and then the constant accidental spectrum. 
A timing window of 30--200 ns is applied to select o-Ps events. 
Figure \ref{fig:energy_spectra} shows the 2$\gamma$ transition spectrum, which is obtained as follows: 
\begin{enumerate}
\item The accidental contribution is subtracted using the timing window $t = $700--900\,ns.
\item The ordinary o-Ps decay spectrum is also subtracted. This spectrum is obtained by running without RF power, 
and is normalized to the 2$\gamma$ transition spectrum in the region 380--460 keV. 
\end{enumerate}
The 2$\gamma$ transition spectrum is measured at different magnetic field strengths. 
The resonance line obtained is shown in Figure \ref{fig:resonance_line}.

Fitting the measured points with the Breit-Wigner function, 
\begin{equation}
\label{bw}
f(B) = \frac{C}{\left( B-B_0 \right) ^2 + \Gamma ^2 /4 } \, ,
\end{equation}
 results in a constant of $C = 2.335(25) \times 10^{-6}$, a center 
value of the magnetic field of $B_0 = 0.865\,6151(99)\,\mathrm{T}$ (11\,ppm), 
and a width of $\Gamma = 3.765(24) \times 10^{-3}\,\mathrm{T}$. 

The main source of systematic error is the uniformity of the magnetic field (142\,ppm in $\Delta _{\mathrm{HFS}}$). 
Other sources are considered to be negligible. 
The preliminary value of $\Delta_{\mathrm{HFS}}$ calculated from equation (\ref{eq:Zeeman}) is 
\begin{eqnarray}
\Delta _{\mathrm{HFS}} &=& 203.399  \nonumber \\ 
 & & \pm 0.005 (\mathrm{stat.}) \pm 0.029 (\mathrm{sys.})\nonumber \\
 & &  \, \mathrm{GHz} \,(143\,\mathrm{ppm})\, ,
\end{eqnarray}
which is consistent with both of the previous experimental values and with the theoretical value.

\subsection{Future plan}
The following improvements are planned for future measurements:
\begin{enumerate}
\item Compensation magnets will be installed and $O(1)$\,ppm magnetic field uniformity is 
expected to be achieved. 
\item Measurements at various pressures of $\mathrm{N}_2$ will be performed to estimate 
the material effect (the Stark effect). The accumulation of these measurements will 
result in an $O(1)$ ppm statistical error within a few years.
\item The timing information allows for a measurement of Ps thermalization as a function of 
time~\cite{kataoka}. We can thus precisely measure the material effect including the 
thermalization effect. 
\end{enumerate}

\section{Conclusion}
A new experiment to measure the Ps-HFS which reduces possible common uncertainties in previous experiments 
has been constructed and the first run is ongoing. A preliminary value of 
$\Delta _{\mathrm{HFS}} = 203.399 \pm 0.005 (\mathrm{stat.}) \pm 0.029 (\mathrm{sys.}) \, \mathrm{GHz} \, (143\,\mathrm{ppm})$ has been obtained, which is 
consistent with both of the previous experimental values and with the theoretical calculation. Development of compensation magnets 
is underway with a view to obtaining $O(1)$\,ppm magnetic field homogeneity for future runs. 
A new result with an accuracy of $O(1)\,\mathrm{ppm}$ will be 
obtained within a few years which will be an independent check of the discrepancy between the present experimental values and the QED prediction.

\end{document}